\documentclass[prl,aps,twocolumn]{revtex4}

\usepackage{psfrag}
\usepackage{graphicx}
\usepackage{dcolumn}
\usepackage{latexsym,amsfonts}
\usepackage{bm}
\usepackage{amssymb}
\begin{document}

\title{On the Time--Modulation of the $\beta^+$--Decay Rate of H--like
  ${^{140}}{\rm Pr}^{58+}$ Ion}

\author{A. N. Ivanov${^{\,a,b}}$, E.  L. Kryshen${^c}$,
M. Pitschmann${^a}$, P. Kienle$^{b,d}$}
\affiliation{${^a}$Atominstitut der \"Osterreichischen
Universit\"aten, Technische Universit\"at Wien, Wiedner Hauptstrasse
8-10, A-1040 Wien, \"Osterreich} \affiliation{${^b}$Stefan Meyer
Institut f\"ur subatomare Physik \"Osterreichische Akademie der
Wissenschaften, Boltzmanngasse 3, A-1090, Wien, \"Osterreich}
\affiliation{${^c}$Petersburg Nuclear Physics Institute, 188300
Gatchina, Orlova roscha 1, Russian Federation}
\affiliation{${^d}$Physik Department, Technische Universit\"at
M\"unchen, D--85748 Garching, Germany} \email{ivanov@kph.tuwien.ac.at}

\date{\today}

\begin{abstract}
 According to recent experimental data at GSI, the rates of the number
  of daughter ions, produced by the nuclear K--shell electron capture
  (EC) decays of the H--like ions ${^{140}}{\rm Pr}^{58+}$ and
  ${^{142}}{\rm Pm}^{60+}$, are modulated in time with periods $T_{EC}
  \simeq 7\,{\rm sec}$ and amplitudes $a_{EC} \simeq 0.20$.  We study
  a possible time--dependence of the nuclear positron $(\beta^+)$
  decay rate of the H--like ${^{140}}{\rm Pr}^{58+}$ ion.  We show
  that the time--dependence of the $\beta^+$--decay rate of the
  H--like ${^{140}}{\rm Pr}^{58+}$ ion as well as any H--like heavy
  ions cannot be observed. This result can be used as a prediction for
  future analysis of the time--dependence of the $\beta^+$--decay
  rates of the H--like heavy ions ${^{140}}{\rm Pr}^{58+}$ and
  ${^{142}}{\rm Pm}^{60+}$ at GSI for the test of the measuring
  method. \\ PACS: 12.15.Ff, 13.15.+g, 23.40.Bw, 26.65.+t
\end{abstract}

\maketitle

\section{Introduction}

The experimental investigation of the $EC$--decays of the H--like ions
${^{140}}{\rm Pr}^{58+}$ and ${^{142}}{\rm Pm}^{60+}$, i.e.
${^{140}}{\rm Pr}^{58+} \to {^{140}}{\rm Ce}^{58+} + \nu$ and
${^{142}}{\rm Pm}^{60+} \to {^{142}}{\rm Nd}^{60+} + \nu$, carried out
at the Experimental Storage Ring (ESR) at GSI in Darmstadt
\cite{GSI2}, showed a modulation in time with periods $T_{EC} \simeq
7\,{\rm s}$ of the rates of the number $N^{EC}_d(t)$ of daughter ions
${^{140}}{\rm Ce}^{58+}$ and ${^{142}}{\rm Nd}^{60+}$, respectively
(see Fig.\,1).  Since the rate of the number of daughter ions is
defined by \cite{GSI2}
\begin{eqnarray}\label{label1}
  \frac{dN^{EC}_d(t)}{dt} = \lambda_{EC}(t)\, N_m(t),
\end{eqnarray}
where $N_m(t)$ is the number of mother ${^{140}}{\rm Pr}^{58+}$ or
${^{142}}{\rm Pm}^{60+}$ ions, a periodic time--dependence can be
attributed to the time--dependent $EC$--decay rate $\lambda_{EC}(t)$
\cite{GSI2}
\begin{eqnarray}\label{label2}
  \hspace{-0.3in}&&\lambda_{EC}(t) = 
\lambda_{EC}\,\Big\{1 + a_{EC}
  \cos\Big(\frac{2\pi t}{T_{EC}} + \phi\Big)\Big\},
\end{eqnarray}
where $\phi$ and $a_{EC}$ are the phase and the amplitude of the
time--dependent term, which are not well--measured quantities
\cite{GSI3,GSI4}. 

In turn, periods of the time--modulation $T_{EC}$ are measured well and
equal to $T_{EC} = 7.06(8)\,{\rm s}$ and $T_{EC} = 7.10(22)\,{\rm s}$
for ${^{140}}{\rm Ce}^{58+}$ and ${^{142}}{\rm Nd}^{60+}$,
respectively \cite{GSI2}. Such a periodic dependence has been
explained in \cite{Ivanov2,Ivanov4,Ivanov5} in terms of the massive
neutrino mixing with the period $T_{EC}$ equal to
\begin{eqnarray}\label{label3}
T_{EC} =  \frac{4\pi \gamma  M_m}{(\Delta m^2_{21})_{\rm GSI}},
\end{eqnarray}
where $M_m$ is the mass of the mother ion \cite{Ivanov5}, $\gamma =
1.43$ is the Lorentz factor \cite{GSI2} and $(\Delta m^2_{21})_{\rm
  GSI} = 2.22(4) \times 10^{-4}\,{\rm eV}^2$, calculated in
\cite{Ivanov2} for the experimental value $T_{EC} = 7\,{\rm s}$
\cite{GSI2}. A relation of the value $(\Delta m^2_{21})_{\rm GSI} =
2.22(4) \times 10^{-4}\,{\rm eV}^2$ to the KamLAND experimental data
$(\Delta m^2_{21})_{\rm KamLAND} = 0.80^{+0.06}_{-0.05}\times
10^{-4}\,{\rm eV^2}$ \cite{PDG06} has been investigated and obtained
in \cite{Ivanov4}. 

The $EC$--decay rate of the H--like ${^{140}}{\rm Pr}^{58+}$ ion,
averaged over time $\langle \lambda_{EC}(t)\rangle = \lambda_{EC}$, as
well as the $\beta^+$--decay rate of the ${^{140}}{\rm Pr}^{58+} \to
{^{140}}{\rm Ce}^{57+} + e^+ + \nu$ decay have been also measured at
GSI \cite{GSI1}: $\lambda^{\exp}_{EC} = 0.00219(6)\,{\rm s^{-1}}$ and
$\lambda^{\exp}_{\beta^+} = 0.00161(10)\,{\rm s^{-1}}$ with the ratio
$R^{\exp}_{EC/\beta^+} = 1.36(9)$.

The calculation of the $EC$ and $\beta^+$ decay rates of the H--like
${^{140}}{\rm Pr}^{58+}$ ion as well as the He--like ${^{140}}{\rm
Pr}^{57+}$ ion has been carried out in \cite{Ivanov1}
\begin{eqnarray}\label{label4}
 \lambda_{EC} &=& \frac{1}{2 F + 1}\,\frac{3}{2} |{\cal M}_{\rm GT}|^2
 |\langle \psi^{(Z)}_{1s}\rangle|^2 \frac{Q^2_{\rm H}}{\pi},\nonumber\\
 \lambda_{\beta^+} &=& \frac{2}{2F + 1}\, \frac{|{\cal M}_{\rm
GT}|^2}{4\pi^3}\, f(Q_{\beta^+},Z - 1),
\end{eqnarray}
where $F = 1/2$ is the total angular momentum of the H--like
${^{140}}{\rm Pr}^{58+}$, $Q_{\rm H} = (3348\pm 6)\,{\rm keV}$ and
$Q_{\beta^+} = (3396 \pm 6)\,{\rm keV}$ are the $Q$--values of the
${^{140}}{\rm Pr}^{58+} \to {^{140}}{\rm Ce}^{58+} + \nu$ and
${^{140}}{\rm Pr}^{58+} \to {^{140}}{\rm Ce}^{57+} + e^+ + \nu$
decays, respectively; 
\begin{figure}[t]
\centering
\includegraphics[width = 0.70\linewidth]{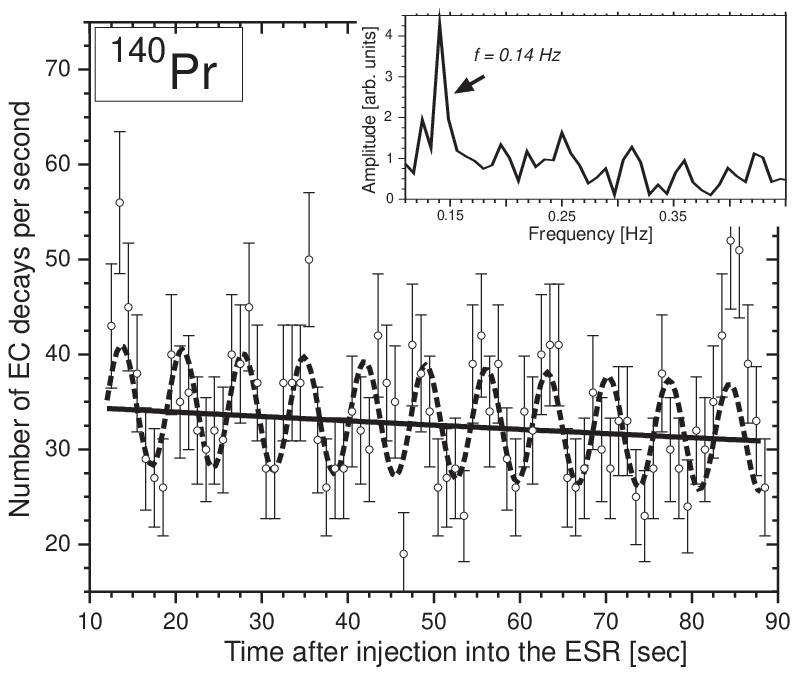}
\includegraphics[width = 0.70\linewidth]{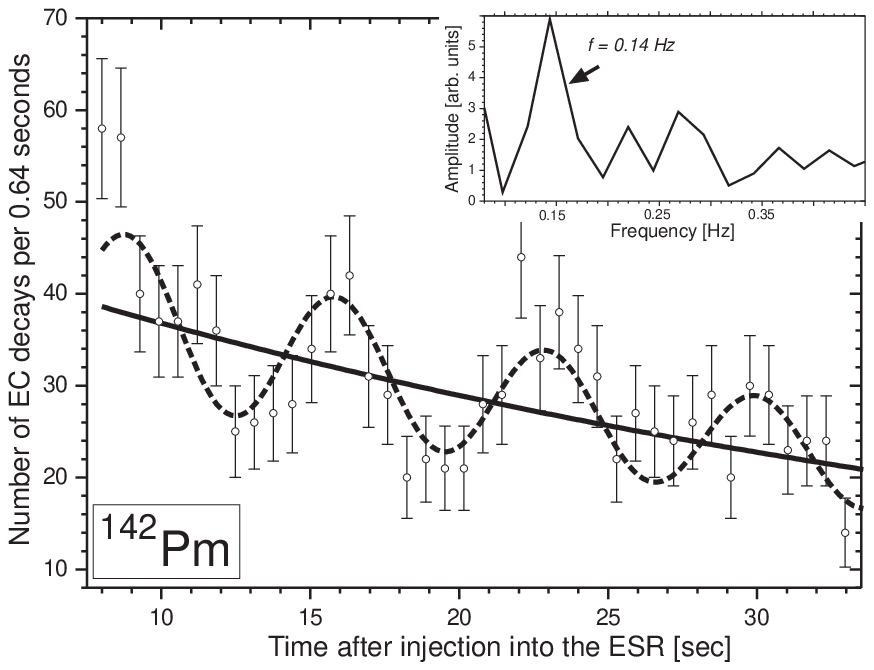}
\caption{The experimental non--exponential time--dependence of the
rate of the number of daughter ions, given by Eq.(\ref{label1}), in
the $EC$--decays ${^{140}}{\rm Pr}^{58+} \to {^{140}}{\rm Ce}^{58+} +
\nu$ and ${^{142}}{\rm Pm}^{60+} \to {^{142}}{\rm Nd}^{60+} + \nu$ of
the H--like ${^{140}}{\rm Pr}^{58+}$ and ${^{142}}{\rm Pm}^{60+}$
ions, respectively.}
\end{figure}
${\cal M}_{\rm GT}$ is the nuclear matrix
element of the Gamow--Teller transition
\begin{eqnarray}\label{label5}
{\cal M}_{\rm GT} = - 2 g_A G_F V_{ud}\int d^3x \Psi^*_d(r)\Psi_m(r),
\end{eqnarray}
where $\Psi^*_d(r)$ and $\Psi_m(r)$ are the wave functions of the
daughter and mother nuclei, respectively, and $\Psi^*_d(r)\Psi_m(r) \sim
\rho(r)$ is the nuclear matter density $\rho(r)$ having the
Woods--Saxon shape with $R = 1.1\,A^{1/3}\,{\rm fm}$ and diffuseness
parameter $a = 0.50\,{\rm fm}$ \cite{NGT}. Then, $\psi^{(Z)}_{1s}$ is
the Dirac wave function of the bound electron in the ground state, $Z
= 59$ is the electric charge of the mother nucleus ${^{140}}{\rm
Pr}^{59+}$.  The average value of the Dirac wave function of the bound
electron $\langle \psi^{(Z)}_{1s}\rangle$ is defined by \cite{Ivanov1}
\begin{eqnarray}\label{label6}
 \langle
  \psi^{(Z)}_{1s}\rangle = \frac{\int\!\! d^3x\psi^{(Z)}_{1s}(\vec{r}\,)
\rho(r)}{
\int\! d^3x\rho(r)} = \frac{1.66}{\sqrt{\pi a^3_B}}
\end{eqnarray}
where $a_B = 1/m_e Z\alpha = 897\,{\rm fm}$ for the electron mass $m_e
= 0.511\,{\rm MeV}$ and the fine--structure constant $\alpha =
1/137.036$. In the $\beta^+$--decay rate $f(Q_{\beta^+},Z - 1) = (2.21
\pm 0.03)\,{\rm MeV}^5$ is the Fermi integral \cite{Ivanov1}. The
theoretical prediction for the ratio $R^{\rm th}_{EC/\beta^+}$ is
\cite{Ivanov1}
\begin{eqnarray}\label{label7}
R^{\rm th}_{EC/\beta^+} = \frac{3 \pi^2 Q^2_{\rm
  H} |\langle \psi^{(Z)}_{1s}\rangle|^2}{f(Q_{\beta^+},Z -1 )} =
  1.40(4),
\end{eqnarray}
which agrees well with the experimental data $R^{\exp}_{EC/\beta^+} =
1.36(9)$.

According to \cite{GSI3,GSI4}, a time--dependence of the rate of the
number of daughter ions in the $\beta^+$--decay of the H--like
${^{140}}{\rm Pr}^{58+}$ ion has not been studied experimentally
until now.

In this paper we apply a theoretical approach, developed in
\cite{Ivanov2}--\cite{Ivanov5} for the analysis of the
time--modulation in the $EC$--decay of the H--like ${^{140}}{\rm
Pr}^{58+}$, to the study of the time--dependence of the
$\beta^+$--decay rate of the H--like ${^{140}}{\rm Pr}^{58+}$ ion. Its
experimental investigation should be a stringent test of the applied
single--ion Schottky mass--measurement method.

\section{Amplitudes of the $\beta^+$--decay of 
the H--like  ${^{140}}{\rm Pr}^{58+}$ ion}

Following \cite{Ivanov2}, for the calculation of the time--modulation
of the $\beta^+$--decay rate of the H--like ${^{140}}{\rm Pr}^{58+}$
ion we use time--dependent perturbation theory. The Hamilton operator
${\rm H}_W(t)$ of the weak interactions is given by
\begin{eqnarray}\label{label8}
 \hspace{-0.3in}&&{\rm H}_W(t ) = \frac{G_F}{\sqrt{2}}V_{ud}\sum_jU_{e
j}\nonumber\\
\hspace{-0.3in}&&\times\int\!\! d^3x [\bar{\psi}_n(x)\gamma^{\mu}(1 -
g_A\gamma^5) \psi_p(x)]\nonumber\\
\hspace{-0.3in}&&\times\,[\bar{\psi}_{\nu_j}(x) \gamma_{\mu}(1 -
\gamma^5)\psi_{e^-}(x)]
\end{eqnarray}
with standard notations \cite{Ivanov2}.  In our analysis neutrinos
$\nu_j\,(j = 1,2,3)$ are Dirac particles with masses $m_j\,(j =
1,2,3)$ \cite{PDG06}.

The amplitude of the $\beta^+$--decay of the H--like ${^{140}}{\rm
Pr}^{58+}$ ion with undetected neutrino we define as a coherent sum of
the amplitudes of the $\beta^+$--decays ${^{140}}{\rm Pr}^{58+}\to
{^{140}}{\rm Ce}^{57+} + e^+ + \nu_j$ \cite{Ivanov2,Ivanov5}
\begin{eqnarray}\label{label9}
\hspace{-0.3in}&& {\cal M}_{FM_F \to F\,' M_{F\,'}}(t) = -
i\sum_j\int^t_{-\infty}d\tau\nonumber\\
\hspace{-0.3in}&&\times\,\langle \nu_j(\vec{k}_j) e^+(p_+)
d(\vec{q}\,))|H_W(\tau)|m(\vec{0}\,)\rangle d\tau,
\end{eqnarray}
where $m$ and $d$ are the mother ion ${^{140}}{\rm Pr}^{58+}$ and the
daughter ion ${^{140}}{\rm Ce}^{57+}$, respectively. The mother ion is
taken in the state ${^{140}}{\rm Pr}^{58+}_{F = \frac{1}{2}}$ with $F
= \frac{1}{2}$ and $M_F = \pm \frac{1}{2}$ and in the rest frame
\cite{Ivanov2}. In turn, the daughter ${^{140}}{\rm Ce}^{57+}$ ion is
a H--like ion in the state with $F\,' = \frac{1}{2}$ and $M_{F\,'} =
\pm \frac{1}{2}$.

The wave function of the neutrino $\nu_j$ we define in the form of a
wave packet \cite{Ivanov2,Ivanov5}
\begin{eqnarray}\label{label10}
\hspace{-0.3in} &&\psi_{\nu_j}(\vec{r},t) = (2\pi \delta^2)^{3/2} \int
\frac{d^3k}{(2\pi)^3}\, e^{\,\textstyle -\,\frac{1}{2}\,\delta^2
(\vec{k} - \vec{k}_j)^2 }\nonumber\\
\hspace{-0.3in} &&\times\,e^{\textstyle\,i\vec{k}\cdot \vec{r} -
iE_j(\vec{k}\,)t}\, u_{\nu_j}(\vec{k},\sigma_j).
\end{eqnarray}
where a spatial smearing of the neutrino $\nu_j$ is determined by the
parameter $\delta$, $\vec{k}_j$ is the neutrino momentum and
$E_j(\vec{k}\,) = \sqrt{\vec{k}^{\,2} + m^2_j}$ is the energy of a
plane wave with the momentum $\vec{k}$,
$u_{\nu_j}(\vec{k},\sigma_{\nu_j})$ is the Dirac bispinor of the
neutrino $\nu_j$ \cite{Ivanov2}.  In the limit $\delta \to \infty$,
due to the relation $ (2\pi
\delta^2)^{3/2}e^{\textstyle\,-\,\frac{1}{2}\,\delta^2(\vec{k} -
\vec{k}_j)^2} \to (2\pi)^3\,\delta^{(3)}(\vec{k} - \vec{k}_j)$, the
wave function (\ref{label10}) reduces to the form of a plane wave
\cite{Ivanov2}.

Following \cite{Ivanov2} and \cite{Ivanov1} we obtain the amplitudes
of the $\beta^+$--decay
\begin{eqnarray}\label{label11} 
  \hspace{-0.3in}&&{\cal M}_{\frac{1}{2},+\frac{1}{2}\to
  \frac{1}{2},M'_F }(t) = -\,\sqrt{2 M_m 2E_d } \frac{{\cal M}_{\rm
  GT}}{2\sqrt{2}}\,(2\pi \delta^2)^{3/2}\nonumber\\
\hspace{-0.3in}&&\times\,e^{\,\varepsilon t}\sum_j
  U_{ej} e^{\,- \frac{1}{2}\,(\vec{k}_d + \vec{p}_+ +
  \vec{k}_j)^2}\,\frac{\displaystyle e^{\textstyle\,i\,\Delta
  E_j(\vec{k}_j)t}}{ \Delta E_j(\vec{k}_j) -
  i\,\varepsilon}\nonumber\\
  \hspace{-0.3in}&&\times\,\Big\{\sqrt{\frac{2}{3}}\,[\varphi^{\dagger}_{n,-\frac{1}{2}}
  \vec{\sigma}\,\varphi_{p,+\frac{1}{2}}]\cdot
  [\bar{u}_{\nu_j}(\vec{k}_j,\sigma_j)\vec{\gamma}\, (1 -
  \gamma^5)\nonumber\\
\hspace{-0.3in}&&\times\,v_{e^+}(\vec{p}_+,\sigma_+)]\,
  \delta_{M'_F,-\frac{1}{2}}- \sqrt{\frac{1}{6}}\nonumber\\
  \hspace{-0.3in}&&\times\, [\varphi^{\dagger}_{n,+\frac{1}{2}}
\vec{\sigma}\,\varphi_{p,+\frac{1}{2}} -
\varphi^{\dagger}_{n,-\frac{1}{2}}
\vec{\sigma}\,\varphi_{p,-\frac{1}{2}}]\cdot
[\bar{u}_{\nu}(\vec{k}_j,\sigma_j)\vec{\gamma}\nonumber\\
\hspace{-0.3in}&&\times\, (1 -
\gamma^5)\,v_{e^+}(\vec{p}_+,\sigma_+)]
\delta_{M'_F,+\frac{1}{2}}\Big\},\nonumber\\
 \hspace{-0.3in} && {\cal M}_{\frac{1}{2},-\frac{1}{2}\to
  \frac{1}{2},M'_F }(t) = -\,\sqrt{2 M_m 2E_d }\, \frac{{\cal M}_{\rm
  GT}}{2\sqrt{2}}\,(2\pi \delta^2)^{3/2}\nonumber\\
\hspace{-0.3in}&&\times\,e^{\,\varepsilon t}\sum_j U_{ej}\,e^{\,-
  \frac{1}{2}\,(\vec{k}_d + \vec{p}_+ +
  \vec{k}_j)^2}\,\frac{\displaystyle e^{\textstyle\,i\,\Delta
  E_j(\vec{k}_j)t}}{ \Delta E_j(\vec{k}_j) - i\,\varepsilon}\nonumber\\
  \hspace{-0.3in}&&\times\,
  \Bigg\{\sqrt{\frac{2}{3}}\,[\varphi^{\dagger}_{n,+\frac{1}{2}}
  \vec{\sigma}\,\varphi_{p,-\frac{1}{2}}]\cdot 
  [\bar{u}_{\nu}(\vec{k}_j,\sigma_j)\vec{\gamma}\,(1 - \gamma^5)\nonumber\\
\hspace{-0.3in}&&\times\,
v_{e^+}(\vec{p}_+,\sigma_+)]\,\delta_{M'_F,+\frac{1}{2}} -
\sqrt{\frac{1}{6}}\nonumber\\
  \hspace{-0.3in}&&\times\,[\varphi^{\dagger}_{n,+\frac{1}{2}}
\vec{\sigma}\,\varphi_{p,+\frac{1}{2}} -
\varphi^{\dagger}_{n,-\frac{1}{2}}
\vec{\sigma}\,\varphi_{p,-\frac{1}{2}}]\cdot
[\bar{u}_{\nu}(\vec{k}_j,\sigma_j)\vec{\gamma}\nonumber\\
  \hspace{-0.3in}&&\times\,(1 - \gamma^5)v_{e^+}(\vec{p}_+,\sigma_+)]
\delta_{M'_F,-\frac{1}{2}} \Bigg\}.
\end{eqnarray}
For the calculation of the amplitudes Eq.(\ref{label11}) we have
 carried out the integration over $\vec{k}$ with the
 $\delta$--function $(2\pi)^3\,\delta^{(3)}(\vec{k} + \vec{k}_d +
 \vec{p}_+)$ and denoted $\Delta E_j(\vec{k}_j) = E_d(\vec{k}_d) +
 E_+(\vec{p}_+) + E_j(\vec{k}_j) - M_m$, where $E_d(\vec{k}_d)$,
 $E_+(\vec{p}_+)$ and $E_j(\vec{k}_j)$ are energies of the daughter
 ion, positron and neutrino $\nu_j$, $\vec{k}_d$, $\vec{p}_+$ and
 $\vec{k}_j$ are their momenta and $M_m$ is the mother ion mass. For
 the calculation of the integral over time we have used the
 $\varepsilon$--regularization \cite{Ivanov2}. Finally the parameter
 $\varepsilon$ should be taken in the limit $\varepsilon \to 0$.

\section{Time--dependence of the  $\beta^+$--decay rate of the 
H--like ${^{140}}{\rm Pr}^{58+}$ ion}

According to \cite{Ivanov2}, the first step to the calculation of the
time--dependent $\beta^+$--decay rate $\lambda^{(\rm H)}_{\beta^+}(t)$
of the H--like ${^{140}}{\rm Pr}^{58+}$ ion is the calculation of the
rate of the neutrino spectrum $N_{\nu}(t)$. It is defined by
\cite{Ivanov2}
\begin{eqnarray}\label{label12} 
\hspace{-0.3in}&& \frac{d N_{\nu}(t)}{dt} = \nonumber\\
\hspace{-0.3in}&& = \frac{1}{2M_m}\,\frac{1}{2 F + 1}\frac{d}{dt}\int
  \sum_{M_F,M'_F}|{\cal M}_{_{F,M_F \to F',M'_F }}(t)|^2\nonumber\\
\hspace{-0.3in}&& \times\,F(Z-1,E_+)\frac{d^3k_d}{(2\pi)^3 2E_d}
  \frac{d^3p_+}{(2\pi)^3 2E_+}
\end{eqnarray}
where $F(Z - 1,E_+)$ is the Fermi function \cite{HS66} (see also
\cite{Ivanov1}), describing the Coulomb repulsion between the positron
and the nucleus ${^{140}}{\rm Ce}^{58+}$. It is equal to \cite{HS66}
\begin{eqnarray}\label{label13}
\hspace{-0.3in}&&F(Z- 1,E_+ ) = \nonumber\\
\hspace{-0.3in}&&=\Big(1 +
\frac{1}{2}\gamma\Big)\,\frac{4(2 R p_+)^{2\gamma}}{\Gamma^2(3 +
2\gamma)}\,e^{\,\textstyle -\,\frac{\pi (Z - 1)\alpha
E_+}{p_+}}\nonumber\\
\hspace{-0.3in}&&\times\,\Big|\Gamma\Big(1 + \gamma - i\,\frac{\alpha (Z - 1)
E_+}{p_+}\Big)\Big|^2,
\end{eqnarray}
where $p_+ = \sqrt{E^2_+ - m^2_e}$, $R = 5.712\,{\rm fm}$, $Z = 59$
and $\gamma = \sqrt{1 - ((Z - 1) \alpha)^2} - 1$.

After the integration over the phase volume of the daughter ion, the
directions of the positron momentum $\vec{p}_+$ and the limit
$\varepsilon \to 0$ we arrive at the following expression for the rate
of the neutrino spectrum
\begin{eqnarray}\label{label14} 
\hspace{-0.3in}&&\frac{d N_{\nu}(t)}{dt} =  \frac{1}{2 F +
1}\,\frac{|{\cal M}_{\rm GT}|^2}{\pi^2}\,(\pi \delta^2)^{3/2}\nonumber\\
\hspace{-0.3in}&&\times\int^{Q_{\beta^+} -
m_e}_{m_e}(2\pi)\delta(Q_{\beta^+} - m_e - E_+ -
E_{\nu})\,E_{\nu}\nonumber\\
\hspace{-0.3in}&& \times\,\Bigg\{1 + \sum_{i > j}2 U^*_{e i}U_{e
j}\,e^{\textstyle - \delta^2 (\Delta
\vec{k}_{ij})^2}\nonumber\\
\hspace{-0.3in}&&\times\,\cos\Big[\Big(\sqrt{E^2_{\nu} + m^2_i} -
\sqrt{E^2_{\nu} + m^2_j}\Big)t\Big]\Bigg\}\nonumber\\
\hspace{-0.3in}&& \times\,F(Z - 1, E_+)\sqrt{E^2_+ - 
m^2_e}\,E_+ dE_+,
\end{eqnarray}
where $\Delta \vec{k}_{ij} = (\vec{k}_i - \vec{k}_j)/2$ and
$e^{\textstyle - \delta^2 (\Delta \vec{k}_{ij})^2}$ are kept as input
parameters \cite{Ivanov2}. 

For the calculation of the r.h.s of Eq.(\ref{label14}) we have set
 neutrino masses zero everywhere except in the energy difference
 $E_i(\vec{k}_i) - E_j(\vec{k}_j)$ \cite{Ivanov2}. Then, due to the
 exponential function $e^{\,\textstyle - \delta^2 ( \frac{\vec{k}_i -
 \vec{k}_j}{2})^2}$ the neutrino momenta are constrained by $\vec{k}_i
 \simeq \vec{k}_j = \vec{k}$ \cite{Ivanov2}. In such an approximation
 we get $E_j(\vec{k}_j) \simeq E_{\nu} = |\vec{k}\,|$ and
 $E_i(\vec{k}_i) - E_j(\vec{k}_j) = \sqrt{E^2_{\nu} + m^2_i} -
 \sqrt{E^2_{\nu} + m^2_j}$. 
\begin{figure}[t]
\centering
\includegraphics[width = 0.80\linewidth]{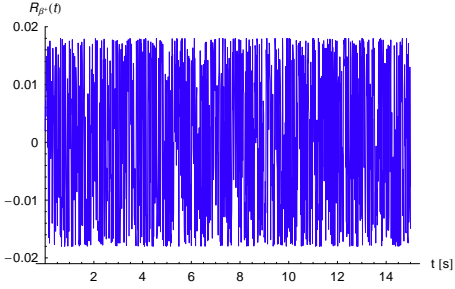}
\caption{Time--dependence of the $\beta^+$--decay rate of the H--like
${^{140}}{\rm Pr}^{58+}$ ion on the time--interval equal to $2 T_{EC} =
14\,{\rm s}$.}
\end{figure}

In terms of the rate of the neutrino spectrum the time--dependent
$\beta^+$--decay rate is defined by \cite{Ivanov2}
\begin{eqnarray}\label{label15} 
\lambda_{\beta^+}(t) = \int \frac{d^3k}{(2\pi)^3 2
E_{\nu}}\,\frac{1}{(\pi \delta^2)^{3/2}}\,\frac{d N_{\nu}(t)}{dt}.
\end{eqnarray}
For $\theta_{13} = 0$ \cite{PDG06} (see also \cite{Ivanov2}) we deal
with two--neutrino mass--eigenstates and obtain
\begin{eqnarray}\label{label16} 
\frac{\lambda_{\beta^+}(t)}{ \lambda_{\beta^+}} = 1 + R_{\beta^+}(t),
\end{eqnarray}
where $\lambda_{\beta^+}$ is given by Eq.(\ref{label4}) and
$R_{\beta^+}(t)$ is equal to
\begin{eqnarray}\label{label17} 
\hspace{-0.3in}&&R_{\beta^+}(t) = \sin2\theta_{12}\,e^{\,\textstyle -
  \delta^2 (\Delta \vec{k}_{12})^2}\int^{Q_{\beta^+} -
  m_e}_{m_e}\!\!\!\!\! dE_+ E_+ \nonumber\\
\hspace{-0.3in}&&\times\,\sqrt{E^2_+ - m^2_e}\,(Q_{\beta^+} -
m_e - E_+)^2\,\frac{F(Z - 1,E_+)}{f(Q_{\beta^+}, Z -
  1)} \nonumber\\
\hspace{-0.3in}&&\times\,\cos\Big[\Big(\sqrt{(Q_{\beta^+} - m_e -
E_+)^2 + m^2_2}\nonumber\\
\hspace{-0.3in}&&- \sqrt{(Q_{\beta^+} - m_e - E_+)^2 +
  m^2_1}\Big)t\Big]. 
\end{eqnarray}
For the numerical calculations we use
  $\sin2\theta_{12}\,\,e^{\,\textstyle - \delta^2 (\Delta
  \vec{k}_{12})^2} = 0.20$ \cite{Ivanov2} and $Q_{\beta^+} =
  3396(6)\,{\rm keV}$ \cite{Ivanov1}.

The time--dependent part of the $\beta^+$--decay rate of the H--like
${^{140}}{\rm Pr}^{58+}$ ion on the time--interval equal to $2 T_{EC}
= 14\,{\rm s}$, i.e. two periods of the time--modulation of the
$EC$--decay rate of the H--like ${^{140}}{\rm Pr}^{58+}$ ion, is shown
in Fig.\,2. For the calculation of $R_{\beta^+}(t)$ we have used
neutrino masses $m_j(R)$, obtained in \cite{Ivanov4} and corrected by
the interaction of massive neutrinos with the strong Coulomb field of
the daughter nucleus ${^{140}}{\rm Ce}^{58+}$ \cite{Ivanov4}. It is
seen that the $\beta^+$--decay rate varies in time much faster than the
$EC$--decay rate of the H--like ${^{140}}{\rm Pr}^{58+}$ ion.

In the measurement of the time--dependence of the $\beta^+$--decay
rate of the H--like ${^{140}}{\rm Pr}^{58+}$ ions the time--spectrum
of the decay is defined by the time--intervals $\Delta T =
5\times\,\Delta T_{\rm bin}$, caused by 5 bins with the length $\Delta
T_{\rm bin} = 64\,{\rm ms}$ each.  This leads to the experimental
value of the $\beta^+$--decay rate, averaged over the time--interval
$\Delta T = 5\times\,\Delta T_{\rm bin} = 0.32\,{\rm s}$. Due to the
rapid variations of $R_{\beta^+}(t)$ a modulation of the
time--dependence of the $\beta^+$--decay rate $\lambda_{\beta^+}(t)$
of the H--like ${^{140}}{\rm Pr}^{58+}$ ion is not observable in an
experiment.

\section{Conclusion}

We have studied the time--dependence of the $\beta^+$--decay rate
$\lambda_{\beta^+}(t)$ of the H--like ${^{140}}{\rm Pr}^{58+}$ ion.
For the calculation of $\lambda_{\beta^+}(t)$ we have followed the
approach, proposed in \cite{Ivanov2}--\cite{Ivanov5} as applied to the
explanation of the time--modulation of the $EC$--decay rate of the
H--like ${^{140}}{\rm Pr}^{58+}$ ion. We have found that the
time--dependent term of the $\beta^+$--decay rate varies very rapidly
in time, which makes such a time--dependence unobservable.  This
result can be used as a prediction for future analysis of the
time--dependence of the $\beta^+$--decay rates of the H--like heavy
ions ${^{140}}{\rm Pr}^{58+}$ and ${^{142}}{\rm Pm}^{60+}$ at GSI for
the test of the measuring method.

\end{document}